\documentclass[reprint]{aastex62}

\graphicspath{{./}{figures/}}
\graphicspath{{./}{Images/}}

\submitjournal{ApJL}

\shorttitle{STORMI Module}
\shortauthors{Roy and Nandy}

\begin{document}

\title{A TIME-EFFICIENT, DATA-DRIVEN MODELLING APPROACH FOR PREDICTING THE GEOMAGNETIC IMPACT OF CORONAL MASS EJECTIONS}

\correspondingauthor{Dibyendu Nandy}
\email{dnandi@iiserkol.ac.in}

\author[0000-0002-0471-4591]{Souvik Roy}
\affil{Center of Excellence in Space Sciences India \\
Indian Institute of Science Education and Research Kolkata \\
Mohanpur 741246, India}

\author[0000-0001-5205-2302]{Dibyendu Nandy}
\affil{Center of Excellence in Space Sciences India \\
Indian Institute of Science Education and Research Kolkata \\
Mohanpur 741246, India}
\affil{Department of Physical Sciences \\
Indian Institute of Science Education and Research Kolkata \\
Mohanpur 741246, India}

\begin{abstract}

To understand the global-scale physical processes behind coronal mass ejection (CME)-driven geomagnetic storms and predict their intensity as a space weather forecasting measure, we develop an interplanetary CME flux rope-magnetosphere interaction module using 3D magnetohydrodynamics. The simulations adequately describe ICME-forced dynamics of the magnetosphere including the imposed magnetotail torsion. These interactions also result in induced currents which is used to calculate the geomagnetic perturbation. Through a suitable calibration, we estimate a proxy of geoeffectiveness -- the Storm Intensity index (STORMI) -- that compares well with the Dst/SYM-H Index. Simulated impacts of two contrasting coronal mass ejections quantified by the STORMI index exhibit a high linear correlation with the corresponding Dst and SYM-H indices. Our approach is relatively simple, has fewer parameters to be fine-tuned, is time-efficient compared to complex fluid-kinetic methods. Furthermore, we demonstrate that flux rope erosion does not significantly affect our results. Thus our method has the potential to significantly extend the time window for predictability -- an outstanding challenge in geospace environment forecasting -- if early predictions of near-Earth CME flux rope structures based on near-Sun observations are available as inputs. This study paves the way for early warnings based on operational predictions of CME-driven geomagnetic storms.

\end{abstract}

\keywords{interplanetary coronal mass ejection, geomagnetic storm, 3D MHD simulation,  geoeffectiveness estimation, space weather prediction}

\section{Introduction} \label{sec:intro} 

Dynamic magnetic conditions on the Sun are known to regulate the near-Earth space environment giving rise to space weather disruptions. The primary drivers of such disturbances are coronal mass ejections (CMEs). Such large-scale solar eruptions originate due to magnetic reconnection, instabilities, and rapid restructuring of coronal magnetic field driven by surface flux emergence and evolution \citep{Yeates_2010, Howard2011}. On their arrival at Earth, their interplanetary counterparts or interplanetary CMEs (ICMEs) \citep{https://doi.org/10.1029/GL009i012p01317, Gopalswamy2006_https://doi.org/10.1007/s11214-006-9102-1} drive geomagnetic storms which can have a significant impact on ground and space-based technologies \citep{SCHRIJVER20152745, Nandy2021}. As we approach the next solar maximum -- predicted to be weak but not insignificant \citep{Bhowmik2018, Nandy_2021} -- we expect to see a rise in the number of such extreme geomagnetic storms. 

Development of prediction techniques for solar flares \citep{Sinha_2022} and CME-induced geomagnetic storms is an outstanding challenge in heliophysics. Usually, the geoeffectiveness of a CME (capability of causing a geomagnetic storm) is quantified using various geomagnetic indices. Of these, the most commonly used are the Disturbance Storm Time Index (Dst, Kyoto Dst) \citep{saguira1991} and SYM-H index \citep{dstsymh}. Although several schemes have been employed to predict the Dst/SYM-H, some or other incompleteness exists in all of these existing paradigms. On one side, the empirical and machine learning model-based predictions that have better efficiency of predicting the storms \citep{https://doi.org/10.1002/swe.20036} entirely lack in describing the global physics of the storm-time geospace. On the other hand, the popular modeling approaches (for example, Space Weather Modelling Framework (SWMF) \citep{https://doi.org/10.1029/2005JA011126}, the Open Geospace General Circulation Model (Open GGCM) \citep{doi:https://doi.org/10.1029/GM125p0377}, and the Coupled Magnetosphere Ionosphere Thermosphere Model (CMIT) \citep{WILTBERGER20041411}, etc.) are not explicitly fined-tuned to estimate these indices but can solve the physics of the interactions. Moreover, in these modeling approaches, kinetic drift-physics models and ionospheric electrodynamics solvers are coupled with a magnetohydrodynamic (MHD) setup -- resulting in computationally expensive calculations. Therefore, to achieve detailed space weather predictions, we need to fill this gap between the understanding of physics and the estimation of geoeffectiveness.  In this study, we demonstrate the possibility of utilizing data-driven MHD modeling of CME flux rope -  magnetosphere interactions towards the purpose. We solve the storm-impacted geospace physics and arrive at a reasonable estimate of geoeffectiveness and temporal variations of geomagnetic storms by taking advantage of fundamental physics.

We develop a 3D MHD STORM Interaction Module (hereby CESSI-STORMI), focusing on a “far-out” planet (like Earth) and the associated magnetosphere. Based on in situ observations at the L1 point, we model incoming transients, namely solar wind and  ICME and introduce these into the domain at the day-side boundary. In this paper, we explore two storms from Solar Cycle 23. The first event (henceforth event 1) occurred on 20 November 2003 which was the strongest geoeffective storm in the last three decades (Dst = $-422\,\mathrm{nT}$) with a highly inclined flux rope with a negative z component of the magnetic field (Bz) in the core \citep{https://doi.org/10.1029/2004GL021639, consuelo2014}.  The second event (henceforth event 2) occurred on 13 April 2006 and was in stark contrast to the previous one; the flux rope of the associated ICME was low in inclination and had a moderate impact on the Earth's magnetosphere (Dst = $-98\,\mathrm{nT}$). While our study uses in-situ data of ICME flux ropes as inputs, we note that if early predictions of such structures become available, our methodology can use these inputs to advance the prediction window.

\section{The STORM Interaction Model Setup} \label{sec:model_setup}

We have developed STORMI based on CESSI-SPIM \citep{sbd2019,10.1093/mnras/stab225} using the magnetohydrodynamic architecture of the open source MHD code PLUTO \citep{pluto2007} (see appendix \ref{A: domain}). In our simulations, the planetary parameters are the same as in \citet{sbd2019} except for the density of the ambient medium, which is remodeled to $5.02 \times 10^{-24} $ gm/cc. The coordinate system is identical to the Geocentric Solar Ecliptic System (GSE) but with a rotation of $180^{\circ}$ about the common z-axis. The initialization of STORMI begins with forcing the planetary dipolar magnetic field with the modeled solar wind and interplanetary magnetic fields (IMFs). The IMF winds from up to 120 minutes before the arrival of the ICME shock are used to establish the magnetospheric steady state based on saturated volume-averaged total energy density inside the computational domain (as defined by \cite{sbd2019} for CESSI-SPIM). We set the magnetic field components of the IMFs by time averaging the corresponding in situ observations of the Wind spacecraft gleaned at Lagrange point L1. The Magnetic Field Instrument (MFI) \citep{Lepping1995_MFI} data from the Wind spacecraft has been obtained from the NASA Coordinated Data Analysis Web (CDAWeb) service (\textit{url: https://cdaweb.gsfc.nasa.gov}).

\begin{figure}[ht!]
\plotone{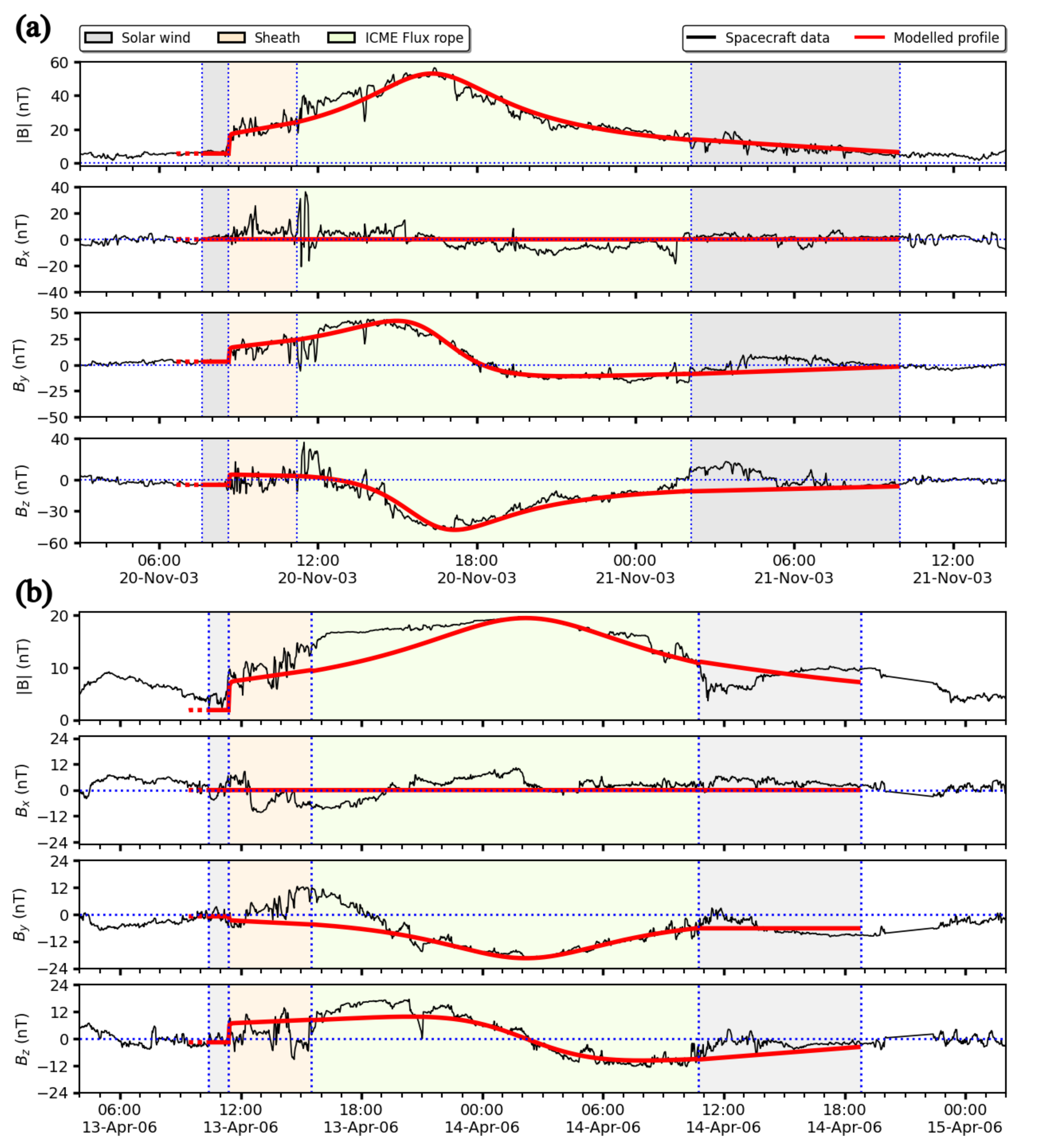}
\caption{Comparison between the ICME structure embedded in the solar wind (black lines) and the modeled input (red lines) in the GSE coordinate system for event 1 in \textbf{(a)} and event 2 in \textbf{(b)}, respectively. The initial dotted red line in the modeled profiles corresponds to the solar wind input that forced the magnetosphere to a steady state. The interplanetary conditions are plotted using the Wind Magnetic Field Investigation (MFI) data provided by the NASA Coordinated Data Analysis Web (CDAWeb) service. The modeled profile is measured at a single point on the x-line at the day side domain boundary and then time-shifted to the average x-position of the Wind spacecraft, which was at the night side of Earth ($\sim\,-213\,\mathrm{R_E}$) for event 1 and at day side ($\sim\,199\,\mathrm{R_E}$) for event 2.}
\label{vir_obs}
\end{figure}

In situ observations show that after the IMF wind, the ICME's shock and sheath region approach Earth, followed by the flux rope \citep{Kilpua2017}. With the realization that a global structure cannot be exactly matched with single-point measurements, we follow simpler, realistic assumptions to model the ICME. The magnetic properties of the sheath regions have been adapted to maintain a smooth transition of the structure from the sheath to the magnetic cloud along the x-axis. We model the flux ropes using the force-free nonlinear flux rope model developed by Gold and Hoyle \citep{10.1093/mnras/120.2.89, hu2014, https://doi.org/10.1002/2016JA023075} which move along the Sun-Earth line (along the x-axis) near 1 AU. The low plasma temperature and low $\mathrm{plasma\, \beta}$ regions in the in situ data determine the temporal boundaries of the modeled flux ropes \citep{https://doi.org/10.1029/JA086iA08p06673}. Although the Gold-Hoyle formalism leads to an axisymmetric magnetic structure of the flux rope, in situ magnetic structure of identified flux ropes does not show radial symmetry. Therefore, we assume that a part of the flux rope is eroded by possible interactions with the interplanetary medium before reaching Lagrange point L1 point such that the remaining flux rope is tapered (see appendix \ref{A: flux rope}). The density and pressure of the ICMEs are defined by solving the Rankine–Hugoniot conditions to generate and maintain the shock fronts. Within the length scale of the domain, which is small compared to the global structure of ICMEs, the central axes of the modeled flux ropes are assumed to be normal to the x-line, and as a result, we get a zero $B_x$ (radial) component inside flux ropes (figure \ref{vir_obs}). The $B_x$ component is taken to be zero for the sheath and solar wind to ease the divergence cleaning process in the simulation. Considering the negligible expansion rate of the cloud, the time-averaged in situ velocities are used as the velocity of the modeled ICMEs. Figures \ref{vir_obs} (a) and (b) show the temporal evolution of the observed magnetic field components of the ambient space environment (black curves) and the modeled wind (red curves) for both events, respectively. Following the ICME transit, we again introduce the IMF forcing on the magnetosphere based on time-averaged data to simulate the relaxation of the magnetospheric system in the aftermath of the storm. Note that while the observations show single-point measurements, the model uses a fitted analytic flux rope, so some differences are unavoidable.
 
\section{Results} \label{sec:results}
\subsection{Magnetospheric dynamics} \label{subsec:dynamics}

In both events, before the ICME enters the domain, the initial solar wind forcing leads to a droplet-shaped, steady, dynamic atmosphere of the planetary magnetosphere \citep{Schwartz1985, sbd2019}. The magnetopause forms on the day side, whereas on the night side, the magnetotail shows the cross-sectional $\theta$ shaped current systems similar to the plasma regimes observed by Geotail \citep{https://doi.org/10.1029/98JA01914} in both cases. However, due to the nonzero y-component of the incoming magnetic field ($B_y$), the $\theta$ boundary assumes an elliptical shape \citep[see][]{https://doi.org/10.1029/97JA00095}. Magnetic flux conservation between the polar cap and the elliptic magnetotail calculated based on the method described by \citet{Kallenrode2001} (chap. 8.2) marks the boundaries of the polar cap around $70.5^{\circ}$ latitude for event 1 and around $73^{\circ}$ latitude for event 2.

\begin{figure}[ht!]
\plotone{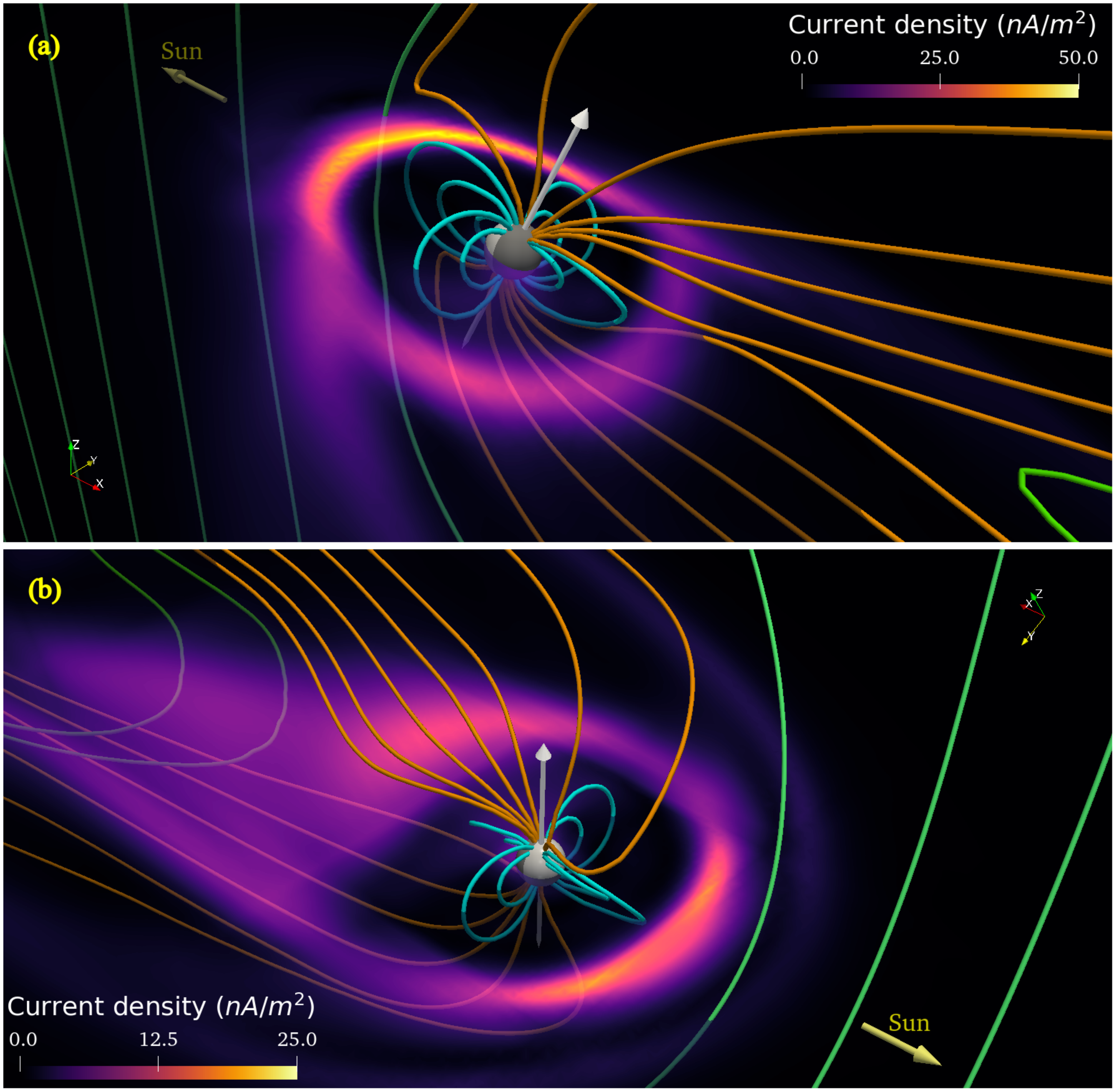}
\caption{Simulated three-dimensional view of the planetary magnetosphere from a viewpoint just above the ecliptic plane. The magnetospheric fields are depicted using colored lines to distinguish among the Earth's polar open field lines (orange), the closed inner magnetospheric lines (cyan), and IMF (green). The strong event (event 1) that occurred on 20 November 2003 is shown in panel \textbf{(a)}, and the moderate event (event 2) of 14 April 2006 is shown in panel \textbf{(b)}. The white arrows in both images denote the rotation axis of Earth. The magnitude of the current density (J) is plotted on the equatorial planes to demonstrate the current formation around the Earth right after the passage of the leading halves of the flux ropes for event 1 (top) and event 2 (bottom). The yellow arrows designate the Sun-side (along the x-axis). 
 }
\label{fig:magnetosphere}
\end{figure}

After the passage of the initial wind, as soon as the ICME shock arrives at Earth, the magnetosphere experiences extreme compression because of the enormous increase in the ram pressure of the inflow. The day-side magnetopause moves toward Earth by a maximum of $2.5 R_E$ for both events. The global magnetosphere undergoes significant perturbations due to the flux ropes passing around the Earth. The planetary magnetic fields change their orientation, influenced by the ram pressure and the magnetic topology of the incoming cloud. Figure \ref{fig:magnetosphere} (a) and (b) shows the 3-dimensional orientation of the magnetosphere and current density at the equatorial planes at 17:33 UTC on 20 November 2003 and at 05:24 UTC on 14 April 2006, respectively, after the leading halves of the modeled flux ropes pass the Earth. The boundary of the simulated polar cap is located around $50^{\circ}$ latitude for event 1 during the maximum impact; in situ measurements confirm that it had moved to about $60^{\circ}$ of geomagnetic latitude (MLAT) in reality \citep{https://doi.org/10.1029/2004JA010924}. On the other hand, for event 2, the boundary of the polar cap shifts to around $55^{\circ}$ of latitude during the impact of the trailing part of the ICME. 

\begin{figure}[ht!]
\plotone{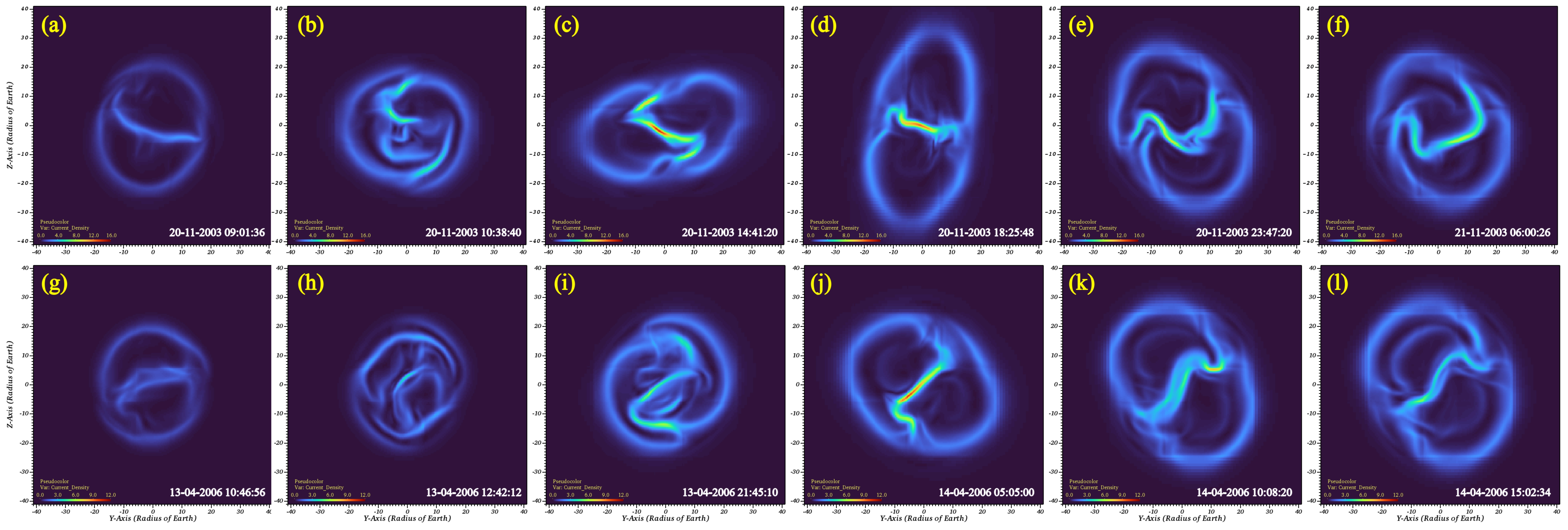}
\caption{Cross-sectional $\theta$-shaped current systems of the magnetotail for event 1 (top row) and event 2 (bottom row). The current density (J) is plotted in the YZ plane (STORMI's coordinate system) at a distance of $60 R_E$ from Earth and is viewed from the magnetotail side. Panels \textbf{(a)} and \textbf{(g)} shows the current boundaries of the magnetotail due to solar wind forcing before the ICME's arrival. In the second column, \textbf{(b)} and \textbf{(h)} reveal the magnetotail currents during the ICME sheath for events 1 and 2, respectively. Panel \textbf{(c)} and \textbf{(i)} show the current system during initial ICME forcing for the events. For the rest of event 1, the magnetotail gets stretched maximally along the semi-major axis as depicted in \textbf{(d)} and starts decreasing in \textbf{(e)} as the trailing part of the ICME crosses the Earth. In \textbf{(f)} the solar wind forcing reshapes the magnetotail away from the perturbed state. On the other hand, \textbf{(j)} and \textbf{(k)} shows the simulated magnetotail during the maximum impact during the transit of the trailing part of the ICME for event 2. Panel \textbf{(l)} shows the post-ICME solar wind-forced magnetotail.}
\label{currentcrosection}
\end{figure}

Studies have shown that the magnetotail retains the memory of the orientation of the previous IMF states since the impact of the solar wind is first observed at the nose of the magnetosphere, followed by dynamic activities of near-Earth magnetic field lines on the night side and then on the distant magnetotail \citep{WALKER1999221, Hultqvist1999}. We observe that the time-varying structure of the magnetic clouds introduces torsion into the magnetotail. Figure \ref{currentcrosection} depicts the simulated time-varying magnetosphere in terms of the cross-sectional current of the magnetotail for the events. During quiet time, the three-dimensional orientation of the geomagnetic dipole governs the orientation of the semi-major axis of the elliptical current system in the magnetotail. However, during the storm, the magnetic forces (and reconnections) acting on the field lines at the polar-cap boundary make the magnetotail cross-section more tilted and elongated because of the flux rope. We notice that the stretching enhances along the semi-major axis with time and becomes maximum when the incoming magnetic field becomes anti-parallel to the dipole axis. Nevertheless, after the passing of the cloud, the quiet time impact of solar wind initiates the relaxation of the storm-forced magnetosphere to the dynamic, steady, quiet time state.

\subsubsection{A New Method for Estimating Geomagnetic Storm Intensity Using the Simulated STORMI Index}

During a storm, disturbances in the magnetosphere are known to be primarily produced by the magnetospheric ring currents, and these disturbances are identified by Dst and SYM-H indices. These two indices measure the storm-associated changes in the magnitude of the axially symmetric component of the geomagnetic field (per hour and minute, respectively), monitored by various low-latitude ground-based observatories \citep[Chap. 8.6]{Menvielle2011}. The currents generated by the perturbed ionosphere and trapped solar wind particles undergo gyro-motion, grad-B drift, and curvature drift motion within the magnetosphere \citep[and references therein]{PhysRev.107.924, Ganushkina2017, Ebihara2019}. Thus, during geomagnetic storms, the enhancement in the entrapped charged particles and the dawn-dusk asymmetry of the magnetic fields increase these currents. This happens in such a way that the geomagnetic field becomes weak, indicated by a reduction in the Dst and SYM-H values.  

In our model, we observe currents in the form $\vec J \propto (\vec\nabla \times \vec B) $ (see appendix \ref{A: domain}) induced by the magnetic field. Images \ref{fig:magnetosphere} (a) and (b) show the distribution of the current density (J) in the equatorial plane of the magnetosphere after the passage of the leading halves of the flux ropes for both events. Although we haven't used any kinetic or particle definition of plasma in our calculation, within the magnetosphere, under the forcing of a southward interplanetary magnetic field (SIMF), these induced currents exhibit a pattern similar to that of the ring currents \citep{10.1002/essoar.10505902.1, 2021EGUGA..23.8863R} when projected on the equatorial plane. The modeled magnetopause boundary current also flows from the west to the east, as seen in \cite{https://doi.org/10.1029/94JA01239}. Moreover, these currents can temporally change their orientations, adapting to the direction of the incoming magnetic field, and we observe a significant increase in their magnitude while the ICME crosses the Earth. The reason behind this behavior of the induced currents is the magnetic topology of the modeled Earth under the influence of the incoming magnetized plasma flow, which can be explained by the principles of fundamental electrodynamics. Based on this understanding, we calculate the STORM Intensity (STORMI) index (as a modeled proxy for the Dst and SYM-H indices) to estimate the geoeffectiveness using the concept of these induced (magnetospheric) currents. To achieve this, we use the integral form of Biot-Savart's law (with usual notations),

\[\vec B(r) = \frac{\mu_0}{4 \pi}\int_{v^\prime} \frac{\vec J(r^\prime) \times \vec r^\prime}{r^{\prime3}} dv^\prime.\]

We assume a disk-like geocentric conductor on the equatorial plane covering the magnetosphere and the plasma atmosphere of the modeled planet with an inner radius of 1.5 $R_E$ (leaving two grid cells from the planetary surface) and an outer radius of 6.5 $R_E$. As the initial current distribution sets the quiet time baseline in our calculation (discussed later), the thickness of the conductor is fixed based on the pre-ICME distribution of the current near the equatorial plane for each event. For event 1, it is 1.9 $R_E$, whereas it is 1.6 $R_E$ for event 2 (see appendix \ref{A: STORMI}). At twelve equatorial grid points (see Figure \ref{equatorialpt} in appendix \ref{A: STORMI}) and two polar grid points just outside the planet, we calculate the expected axial component (parallel to the dipole axis) of the magnetic field for the current distribution throughout the volume ($v^\prime$) of the plasma sphere. We take the mean contribution from all fourteen points to calculate the global induced field. \\

\begin{figure}[ht!]
\plotone{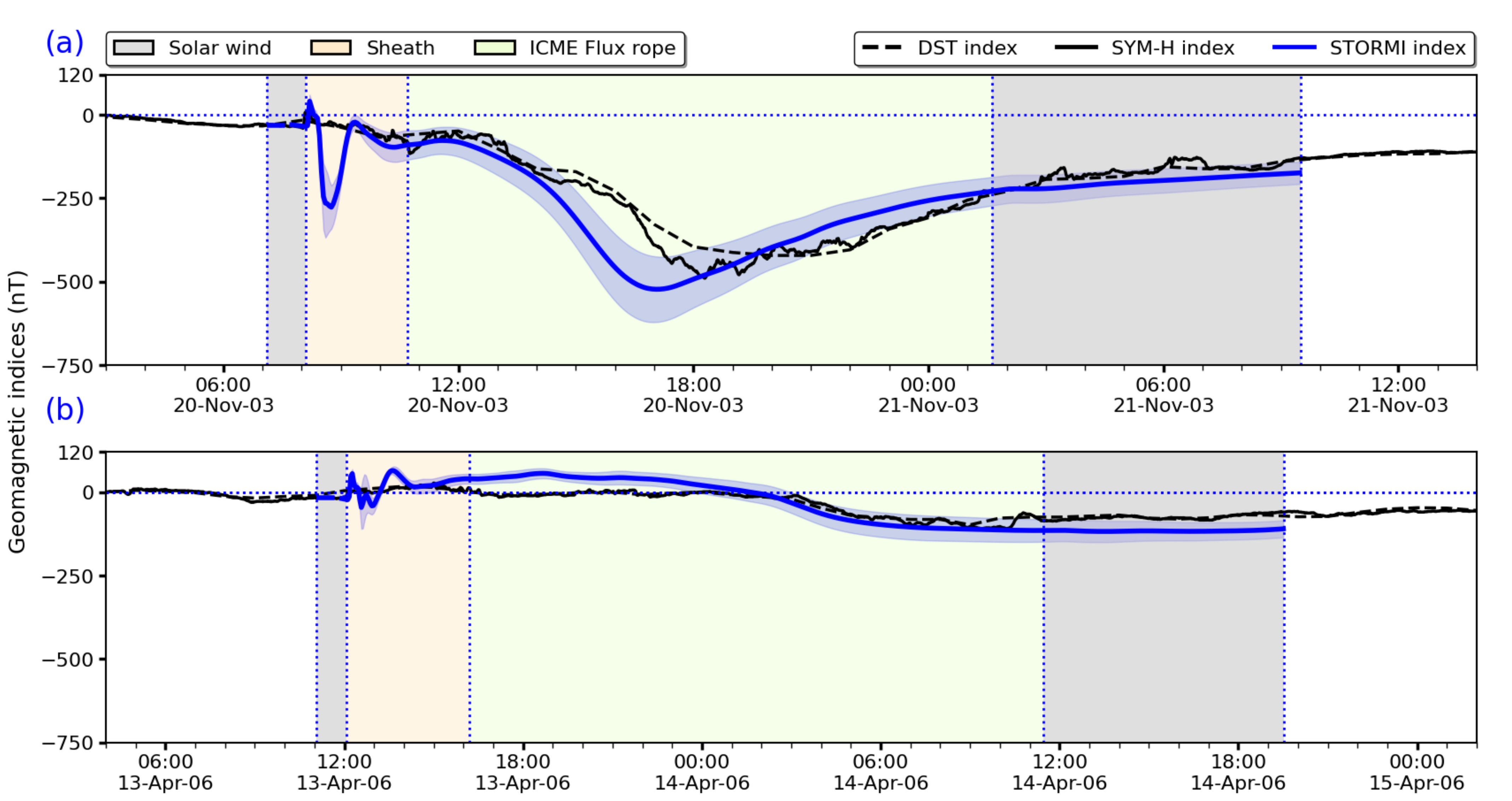}

\caption{The figure presents the comparison of the time evolution of the Dst index (black dashed curve), SYM-H index (solid black line), and the modeled reduction in the geomagnetic field intensity in terms of STORMI index (blue curve) for event 1 in panel \textbf{(a)} and for event 2 in panel \textbf{(b)}. The semi-transparent blue-shaded region is the standard deviation of the STORMI geomagnetic storm index. The STORMI index is computed after incorporating the time delay between the in situ observations of the solar wind (shown in figure \ref{vir_obs}) and its geomagnetic perturbation.}
\label{inducedmagcur}
\end{figure}

As previously mentioned, we consider the current distribution during the pre-ICME solar wind forcing as the reference condition of the magnetosphere in calculating the STORMI index. Accordingly, we designate the corresponding field as the quiet-time baseline and estimate the STORMI index by calculating the change in the induced magnetic field. In figure \ref{inducedmagcur} (a) and (b), the blue curve shows the progression of the STORMI index with time for both events, along with the shaded blue regions showing the span of the standard deviation ($\sigma$). The time axes in figure \ref{inducedmagcur} are calculated after incorporating the time taken by the respective flux ropes to reach Earth from its in situ observation point. The index for event 1 reaches the maximum value (with $\sigma$ bound) of $-523.9\pm98\,\mathrm{nT}$ at 17:03 UTC on 20 November 2003, as the flux rope's negative $B_z$ core reaches the Earth. For event 2, the index remains steady with time due to induced eastward currents resulting from the positive $B_z$ of the incoming flow. However, as soon as $B_z$ changes polarity for event 2, we see a rapid fall in the index, which reaches a moderately low value of $-114.6\pm34\,\mathrm{nT}$ at 11:19 UTC on 14 April 2006. Moreover, in both cases, we see a slow recovery of the index as the solar wind replaces the ICME and the system relaxes. 

As expected, these results are sensitive to the modeled analytical profile. A 10$\%$ increase in the axial magnetic field ($B_0$, refer to appendix \ref{A: flux rope}) of the modeled flux rope leads to around a 6.4$\%$ increase in the magnitude of the STORMI index, which closely matches the estimate by \citet{Grayver_2022}. Also, a 10$\%$ change in the twist of the modeled flux rope changes the index by only 1.3$\%$. These, nonetheless, underscore the importance of fitting the flux rope well. For qualitative comparisons, we plot the real-time values of Dst and SYM-H in black dashed and solid curves in figure \ref{inducedmagcur}. The Dst and SYM-H data has been acquired from the Geomagnetic Data Service by World Data Center for Geomagnetism, Kyoto (\textit{url: https://wdc.kugi.kyoto-u.ac.jp/wdc/Sec3.html}). The minimum Dst value for event 1 was $-422\,\mathrm{nT}$, recorded at 20:00 UTC, whereas the minimum SYM-H was $-490\,\mathrm{nT}$, recorded at 18:17 UTC. The Dst and SYM-H values for event 2 were $-98\,\mathrm{nT}$ and $-111\,\mathrm{nT}$ recorded at 09:00 UTC and 09:21 UTC on 14 April 2006, respectively. For the quantitative analysis of the temporal variation between the STORMI index with the Dst and SYM-H indices, we calculate the Pearson linear correlation coefficient (r). For both events, the observed and modeled indices exhibit a linear relationship between the STORMI index and Dst and SYM-H. The Pearson coefficients are 0.83 and 0.88 for Dst and SYM-H prediction, respectively, for event 1. The same is 0.95 and 0.93, respectively, for event 2.

\section{Discussion and Conclusions} \label{sec:discussion}

We develop a 3D MHD module called CESSI-STORMI to simulate the impacts of ICMEs on planetary magnetospheres. Our approach is physically intuitive, has fewer parameters to be fine-tuned as compared to more complex models, and has the potential to significantly extend the time window for predictability – which is one of the most critical factors in geospace environment prediction. In our data-driven simulations, instead of mapping a single point observation from Lagrange point L1 to a 2D (in-flowing) boundary like other physics-based modules, we use Gold-Hoyle-type magnetic flux ropes to model the 3D ICME structure. This approach produces a better physical representation of the global magnetic structure of the storm Also, we introduce the simulated STORMI index to quantify the geo-effectiveness of the events as a proxy for the Dst and SYM-H index. STORMI index estimates the reduced magnetic field -- averaged over fourteen different points around the modeled Earth --- due to induced currents in the equatorial region of the simulated magnetosphere. 

Like the available physics-based models that estimate Dst/SYM-H, we utilize Biot-Savart's law for calculating the STORMI index. This method is much more efficient and less erroneous than calculating the index by measuring the change in the magnetic field at the grid points. The expected change in these points is roughly one-thousandth of their magnetic field value, and any small change due to advection would be misleading. Also, our approach is simpler because we do not incorporate any additive complexities from coupling the electrodynamics solvers. It is also evident that the STORMI index, calculated using simulated equatorial induced currents, shows high efficiency in predicting the strength of a storm as well as its temporal variation and the quality of modeling of the incoming flux ropes plays an important role in this achievement. Our calculations for two contrasting (strong and moderate) geomagnetic storm events demonstrate a very good correlation (Pearson's coefficient greater than 0.8) between observed and simulated geomagnetic indices. This performance is comparable with, if not better than the available physics-based and empirical/machine learning model-based predictions \citep{https://doi.org/10.1029/2018SW002067}. And with this capability, our paradigm bridges a crucial gap between empirical and computationally heavy, coupled MHD-electrodynamic solvers. 

Draping of the interplanetary medium around a propagating flux rope may change the structure of a propagating CME in the heliosphere \citep{https://doi.org/10.1029/2018SW001944, Pal_2020}. Thus in-situ observations at L1 are currently relied upon for flux rope reconstructions. Moreover, inputs derived from L1 to simulate geoeffectiveness do not leave enough time-window for forecasting a storm. Therefore, the current utilization of STORMI is limited to post-diction analysis of geomagnetic storms. However, with advances in predicting the properties of ICME magnetic clouds near Earth based on near Sun observations, early knowledge of the properties of the 3D magnetic structure, speed, and passage time of flux ropes may now be possible \citep{Pal_2018, Pal_2022}. This suggests that the STORMI module can be used at an earlier phase, up to 1--2 days in advance, for forecasting the timing and intensity of geomagnetic storms.

To assess this possibility, we have performed a comprehensive analysis of the impact of flux rope erosion and tapering on the simulated storm intensity index (see appendix \ref{A: flux rope}). Our simulations indicate that even if the boundary regions of flux rope structures undergo erosion and consequently become tapered, the impact on the simulated minimum of the storm intensity index (at the time of maximum storm impact) is negligible (see appendix \ref{A: flux rope}). Therefore, with available inputs of early predictions of near-Earth CME flux rope structures, our methodology can become a transformative tool in timely forecasts of CME induced geomagnetic storm intensities.

{\acknowledgments The Center of Excellence in Space Sciences India (CESSI) is funded by the Ministry of Education, Government of India. Souvik Roy acknowledges the financial support from the Human Resource Development Group (HRDG) of the Council of Scientific and Industrial Research (CSIR), India. The authors acknowledge Bhargav Vaidya, Sanchita Pal, Srijan B. Das, Arnab Basak, and Yoshita Baruah for useful discussions. We are grateful to Prosenjit Lahiri for his contributions to maintaining the computational infrastructure at CESSI. We are thankful to the anonymous reviewer for their insightful feedback.}

\vspace{5mm}

\appendix

\section{Description of Computational Domain} \label{A: domain}

Similar to CESSI SPIM \citep{sbd2019,10.1093/mnras/stab225}, in CESSI-STORMI, we simulate a 3-dimensional domain developed in PLUTO \citep{pluto2007} architecture considering the interplanetary space, the planetary atmosphere and the solar wind as a single-fluid plasma. The interactions are governed by the adiabatic equation of state and the following set of resistive MHD equations.

\[ \frac{\partial \rho}{\partial t} + \nabla \cdot (\rho \mathbf{v})  =  0 \]
\[   \frac{\partial (\rho  \mathbf{v})}{\partial t} + \nabla \cdot \left[\rho \mathbf{v}\mathbf{v} - \mathbf{B}\mathbf{B}\right] + \nabla \left( p + \frac{\mathbf{B}^2}{2} \right)  = \rho \mathbf{g}  \]
\[   \frac{\partial E_{t}}{\partial t} + \nabla \cdot \left[ \left( \frac{\rho\mathbf{v}^2}{2} +\rho e + p\right)\mathbf{v}  + c\mathbf{E}\times \mathbf{B} \right] = \rho  \mathbf{v} \cdot \mathbf{g}  \]
\[   \frac{\partial \mathbf{B}}{\partial t} + \nabla \times (c\mathbf{E}) = 0. \]

The variables $\rho$, $\mathbf{v}$, $\mathbf{B}$, $p$, $E_t$ denote the density, velocity, magnetic field, pressure, and total energy density, respectively. A factor of $\frac{1}{\sqrt{4\pi}}$ has been incorporated in the definition of $\mathbf{B}$ such that the total energy density $E_{t}$ for an ideal gas can be written as,

\[ E_{t} = \frac{p}{\gamma-1} + \frac{\rho \mathbf{v}^{2}}{2} + \frac{\mathbf{B}^2}{2} \]

We use $\mathbf{g}$ as the acceleration experienced by the fluid due to the gravitational field of the planet in terms of the body force vector. The electric field, \textbf{E}, having a convective and a resistive component, is written as,

\[ c\mathbf{E} = -\mathbf{v}\times \mathbf{B} + \frac{\eta}{c} \mathbf{J}. \]

Here $\mathbf{J}= c\,\nabla\times\mathbf{B}$ is the current density. We have neglected displacement currents in the system. Also, we consider a finite and isotropic magnetic diffusivity $\mathbf{\eta}$ which has a constant value ($10^{9} \, m^2 s^{-1}$) throughout the simulation as the causal mechanism for non-ideal processes such as magnetic reconnections \citep{https://doi.org/10.1029/1999JA900159, 1996ESASP.389..459W}.

To perform the integration, we have used HLL Riemann solver and linear interpolation in space, MINMOD limiter, and 2$^{\rm nd}$ order Runge-Kutta with ``super-time-stepping" for the temporal update. We impose the $\nabla \cdot \mathbf{B} = 0$ condition using the divergence-cleaning method, an approach based on the generalized formulation of the Lagrange multiplier (GLM). The left boundary of the domain is used as an inflow boundary for the solar wind and ICME, whereas the other five are ``outflow" boundaries for all the parameters.

In all three directions, the computational domain extends from $-205 R_{\rm E}$ to $205 R_{\rm E}$. A zone-wise static mesh refinement is implemented in the directions of the 3D Cartesian grid configuration such that the distance from the center to $6 R_{\rm E}$ in all directions has been resolved with a grid-size of $0.3 R_{\rm E}$. From $6 R_{\rm E}$ to $25 R_{\rm E}$ the resolution is $0.5 R_{\rm E}$. The length of the grid-size increases to $1 R_{\rm E}$ between $25 \sim 50 R_{\rm E}$, then to $2 R_{\rm E}$ between $50 \sim 100 R_{\rm E}$ and finally to $3 R_{\rm E}$ in between $100 \sim 205 R_{\rm E}$.

Although Earth's magnetic axis is tilted by an angle of about $11^{\circ}$ from the rotation axis, since, for this study, the computational timescale is longer than a day and the modeled Earth is stationary, we consider the rotation axis to be the best possible time-averaged magnetic axis such that the geographic and magnetic equators become coplanar. Due to the time independence of the planetary field, we can adapt a computationally convenient formalism in solving the coupled MHD equations that deal with high-beta plasma \citep{Powell1997}. We treat the total magnetic field inside the computational domain as a sum of curl-free, time-invariant, background magnetic field (contributed by the dipole) and a deviation due to forced plasma interactions such that the energy depends on the deviated field only.

\section{Description of Flux Rope Modelling} \label{A: flux rope}

We model the flux ropes using the Gold-Hoyle (GH) tube \citep{10.1093/mnras/120.2.89}, where the twist is radially uniform, the helicity is positive, and the magnetic field components, in cylindrical coordinates ($x = r cos \phi, y = r sin \phi, z = z $), are,

\[B_r = 0, \]
\[B_{\phi} = \frac{Tr}{1+T^2r^2}B_0, \]
\[B_{z} = \frac{1}{1+T^2r^2}B_0. \]

Here $B_0$ is the magnitude of the magnetic field at the axis of the flux rope, and T is a constant representing the twist per unit length in the form of $\tau = \frac{T}{2\pi}$. 

\begin{figure}[ht!]
\plotone{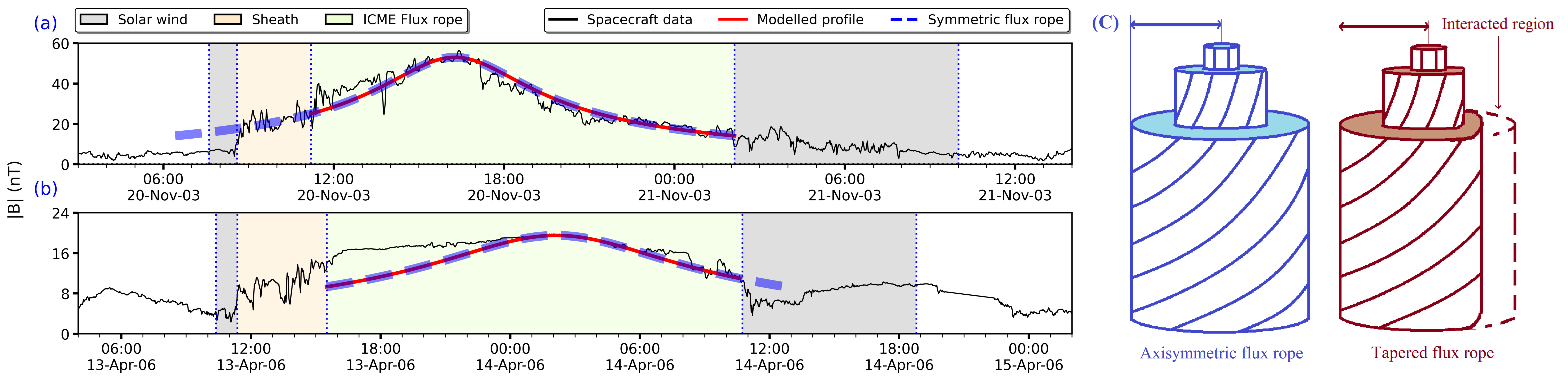}
\caption{In the left column, the in situ profiles of the magnitude of the magnetic field for both events are plotted in black in panels \textbf{(a)} and \textbf{(b)}, respectively. The background in green signifies the flux rope region that shows a drop in plasma $\beta$ value. We consider a tapered flux rope model to develop the 3D structure of the ICMEs, as shown in the red lines. The dashed blue lines in both panels depict the associated axially symmetric flux rope near Earth without any interaction during the heliospheric propagation of the CME. In reality, some part of that symmetric flux rope (the leading edge for event 1 in panel \textbf{(a)} and the trailing edge for event 2 in panel \textbf{(b)}) vanishes due to possible interactions with the interplanetary medium before reaching Earth. The remaining modeled flux rope is shown using the red line within the green patched region. In panel \textbf{(c)}, we plot a cartoon of an axisymmetric flux rope (in blue) and a tapered flux rope (in maroon) side by side for an overall understanding. The tapered flux rope has an eroded edge due to interplanetary interactions. The radius of the axisymmetric flux rope is taken to be the same as the length between the core and the distant boundary of the tapered flux rope (marked with arrows).}
\label{tapered_fr}
\end{figure}

The Gold-Hoyle formulations show that the flux rope will have an axisymmetric structure with a maximum magnetic field magnitude in its core that decreases with increasing radius. Near-Earth flux ropes can be viewed as axisymmetric structures in an ideal scenario where interplanetary structures are interacting while propagation.  However, in reality, when a flux rope approaches Earth, interplanetary interactions come into play and cause some parts of the flux rope to erode, leading to an asymmetry in the magnetic field amplitude (as shown in the green patched regions of the panel (a) and (b) in figure \ref{tapered_fr}). Thereby to facilitate the modeling process, we assume that we observe a tapered flux rope near Earth that shows deviations from an axisymmetric structure due to interplanetary conditions. In simpler terms, the tapered flux rope can be considered as an axisymmetric flux rope which, instead of reaching Earth with an intact structure, interacts with the interplanetary medium while propagating and is eroded (refer to the artistic impression in figure \ref{tapered_fr}(c) for a better understanding). Thus, it loses its symmetry, as observed in the single-point measurements.

To reconstruct the 3-dimensional structure of the tapered flux rope using Gold Hoyle equations, we start with modeling the associated axisymmetric structure. For the sake of simplicity, we consider that only one edge (either leading or trailing) of the axisymmetric tube erodes. As a result, the radius of the axisymmetric rope remains the same as the spatial distance between the maximum magnetic field point and the distant boundary of the tapered structure, as seen in figure \ref{tapered_fr}(c). We model fit the in situ observation data with the associated Gold Hoyle flux rope within the low plasma $\beta$ region (green patched regions in the panel (a) and (b) in figure \ref{tapered_fr}) and input this un-impacted flux rope in our simulation. Following standard protocol, initiating with guess values of the axial magnetic field, twist, and orientation of the flux rope and we perform the least squared fitting throughout individual parameter spaces to find the best-fit values. For event 1, we get $B_0$ to be $-56 \, \mathrm{nT}$ and $\tau$ to be $+3.8$ per astronomical unit (AU) as the best-fit parameters. The flux rope axis is inclined by an angle of $-54^o$ to the ecliptic plane. On the other hand, we get $B_0$ is $-19.5\,\mathrm{nT}$ and $\tau$ is $-2.2$ per AU for event 2, where the flux rope axis is parallel to the ecliptic plane. In the left panels of figure \ref{tapered_fr}, the red line shows the model-fitted structure of the magnitude of the magnetic field within the identified region of flux rope (green patched), which we input in our simulations in terms of x, y, z components of magnetic field (figure \ref{vir_obs}). In the same panels, the blue dashed line depicts the associated axisymmetric flux rope structures for both events.

\begin{figure}[ht!]
\plotone{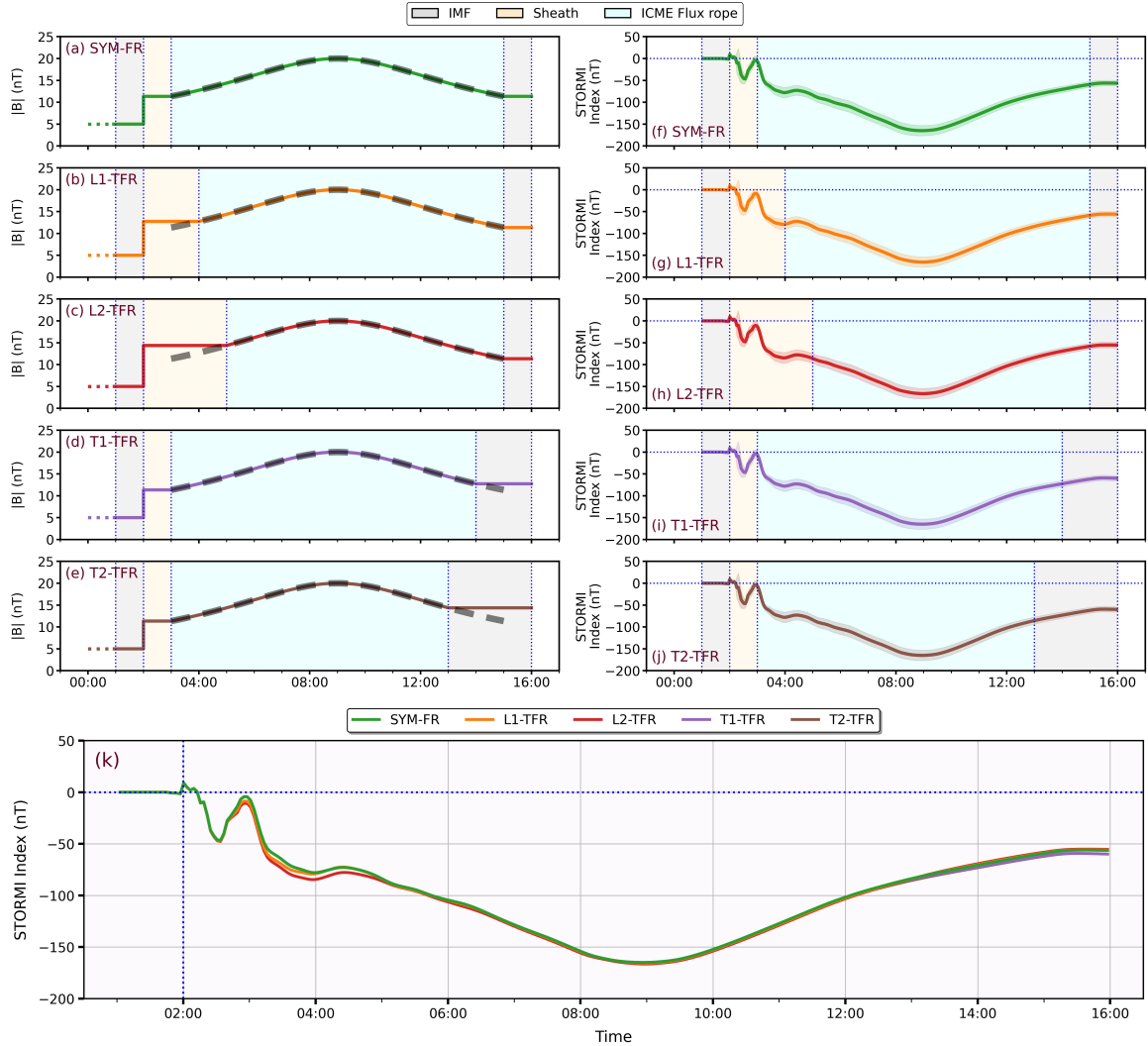}
\caption{The top panels (\textbf{a}-\textbf{j}) show side-by-side comparisons of modeled magnetic field profiles (left column) and corresponding STORMI (storm intensity) index profiles (right column) for different erosion scenarios. The standard deviations for the estimates are encapsulated in the width of the STORMI index curves. Each row represents a different case which is noted in the plots. The black dashed lines in the plots on the left column are fits assuming an underlying axisymmetric profile; these indicate departures from an axisymmetric flux rope (as in figure \ref{tapered_fr}). Timestamp is maintained at Earth's location to account for propagation delays. The bottom image (panel \textbf{k}) displays a superimposed plot of the STORMI index for all cases. The maximum change in the index due to tapering is approximately $9\,\mathrm{nT}$ which is in the leading edge regions, while in the trailing edge regions, it is less than $3\,\mathrm{nT}$. However, these changes are localized near the flux rope boundary for individual cases and do not significantly impact the overall storm intensity profiles or the maximal impact.}
\label{index_tapered}
\end{figure}

We investigate the impact of flux erosion and tapering on the STORMI simulated storm intensity index by simulating diverse scenarios of a general, inclined axisymmetric flux rope with different levels of erosion. We model five cases: (a) no erosion (SYM-FR), (b) leading edge eroded by one hour (L1-TFR), (c) leading edge eroded by two hours (L2-TFR), (d) trailing edge eroded by one hour (T1-TFR), and (e) trailing edge eroded by two hours (T2-TFR). All other properties of the flux ropes remain unchanged. We use specific values for initial IMF ($0,\,0,\,-5\,\mathrm{nT}$), solar wind, and ICME velocity ($400\,\mathrm{km/s}$ and $800\,\mathrm{km/s}$, respectively), axial magnetic field ($B_0\,=\,-20 \,\mathrm{nT}$), and twist ($\tau\,=\, +2.0/\mathrm{AU}$) and obtain the storm intensity index values during maximum impacts to be $-164.98\,\pm\,11.60\,\mathrm{nT}$ (for SYM-FR, T1-TFR, and T2-TFR), $-165.46\,\pm\,11.62\,\mathrm{nT}$ (for L1-TFR), and $-165.57\,\pm\,11.67\,\mathrm{nT}$ (for L2-TFR). This indicates that the maximum change in the minimum index (at time of maximum storm impact) for tapering is less than 0.4\% (for the leading edge erosion); no changes are observed for trailing edge erosions as expected at time of maximum impact. We display the flux rope profiles and corresponding STORMI index profiles in the top columns of figure \ref{index_tapered} and present a visual summary of the results in the bottom panel of figure \ref{index_tapered}, indicating that differences in the storm intensity profiles are not significant. These findings suggest that the storm intensity index is not critically dependent on the tapering of flux rope boundaries. We note additionally that our findings for these theoretically simulated flux ropes closely match the statistically estimated Dst value reported by \citet{KANE2010392} (see Table 3). This lends further credence to the reliability of the STORMI index as a modeled proxy for Dst.

\section{Details of STORMI Index calculation} \label{A: STORMI}

\begin{figure}
\plotone{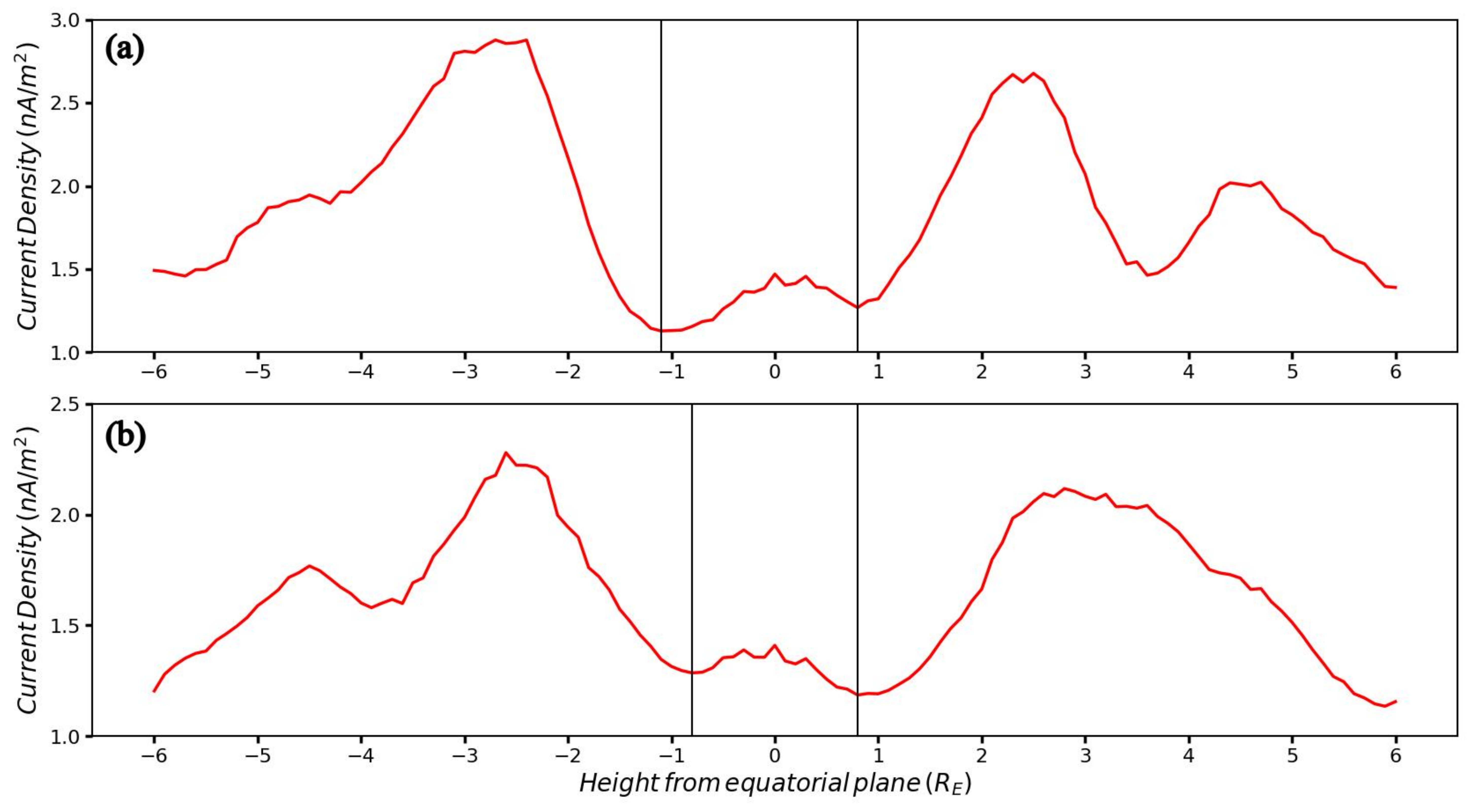}
\caption{Panel \textbf{(a)} and \textbf{(b)} depicts the distribution of time-averaged quiet time current density, surface-averaged over planes parallel to the equatorial plane for event 1 and event 2, respectively, as a function of the height from the equator. The center of the horizontal axis denotes the equator with zero height. A positive height represents a plane in the northern hemisphere, and a negative height indicates a plane in the southern hemisphere. The vertical black lines mark the height of the plane that has a local minimum value of average current near the equator.}
\label{currentavg}
\end{figure}

To estimate the width of the current distribution near the equator during quiet time, we consider a cylindrical tube in the domain with an inner radius of 1.5 $R_E$ and an outer radius of 6.5 $R_E$. The axis of the tube is parallel to Earth's rotation axis, and the height ranges from $\mathrm{-6}$ to 6 $R_E$, keeping Earth at the center. Throughout the simulated quiet time, we surface-average the current density over circular cross-sections of the cylinder parallel to the equator and plot the time-averaged value as a function of their distance from the center of the Earth in figure \ref{currentavg}. The width of the circular planes is taken to be $0.1 R_E$. 

\begin{figure}[ht!]
\plotone{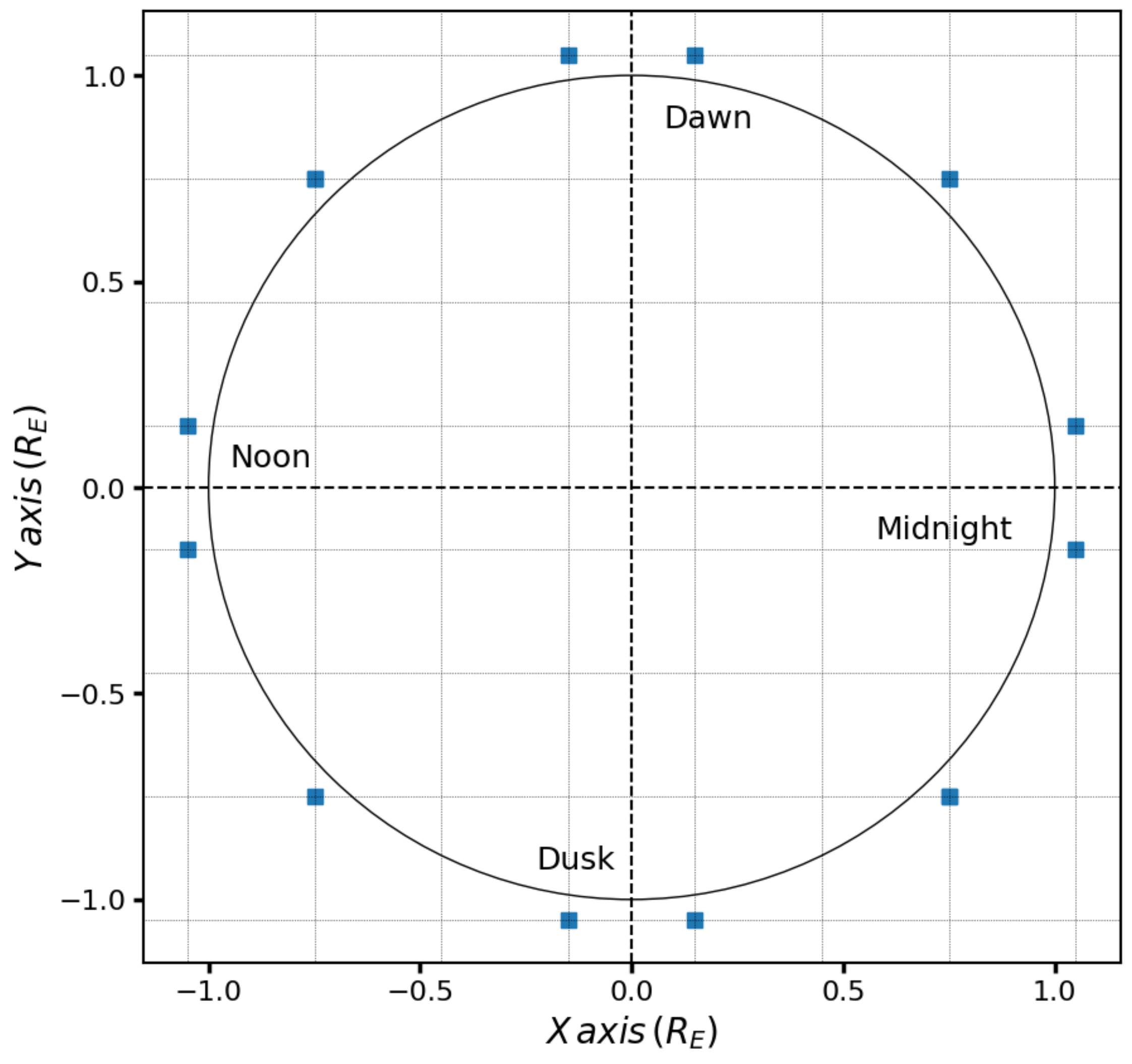}
\caption{The figure shows the 2-dimensional distribution of the twelve points around the equator marked as blue squares. The equator is projected to the XY-plane in STORMI's coordinate system. The points are chosen as the nearest grid centers placed just outside the black circle with a radius equal to Earth's. The background cells correspond to the near-Earth grid resolution of STORMI, where each corner represents a grid center.}
\label{equatorialpt}
\end{figure}

The current shows a local maximum in the distribution near zero (equator), indicating the equatorial accumulation of the simulated currents. However, as we move to higher latitudes, the current increases rapidly after falling to a dip. These high currents near $\pm$ 3 $R_E$ (in figure \ref{currentavg}) arise from the contributions of the polar cap regions above the poles. We do not consider these currents in calculating the STORMI index and estimate the width of the distribution within the local minimum values near the zero point (the equator), identified with black vertical lines. Hence we consider the width to be 1.9 $R_E$ for event 1 and 1.6 $R_E$ for event 2. We calculate the expected axial component of the magnetic field due to the current distribution at twelve equatorial grid points (as depicted in figure \ref{equatorialpt}) and two polar grid points as virtual observatories on the surface of the modeled Earth. The mean contribution from all fourteen points results in the STORMI index.

\newpage
\newpage
\bibliographystyle{apalike}

\bibliography{manuscript.bib}

\newpage
\newpage

\end{document}